\documentstyle[prl,aps,epsf]{revtex}
\begin{document}
\draft
\title{ Unique Spin Dynamics and Unconventional Superconductivity in the Layered Heavy Fermion Compound CeIrIn$_5$: NQR Evidence}
\author{G.-q. Zheng$^1$, K. Tanabe$^1$, T. Mito$^1$, S. Kawasaki$^1$,  Y. Kitaoka$^1$, D. Aoki$^2$, Y. Haga$^3$, and Y. Onuki$^2$}
\address{$^1$Department of Physical Science, Graduate School of Engineering
Science, Osaka University,  Osaka 560-8531,
Japan}
\address{$^2$Department of Physics, Graduate School of  Science, Osaka
University, Toyonaka, Osaka 560-0043,
Japan}
\address{$^3$Advanced Science Research Center, Japan Atomic Energy Research Institute, Tokai, Ibaraki 319-1195, Japan.}
%\date{received  Nov.1, 2000}
\twocolumn[
\maketitle
\widetext
\begin{abstract}
 We report measurements of the $^{115}$In  nuclear spin-lattice relaxation rate ($1/T_1$)  between $T$=0.09 K and 100 K in the new heavy fermion (HF) compound CeIrIn$_5$. At 0.4 K $\leq T \leq$ 100 K, $1/T_1$ is strongly $T$-dependent, which indicates that CeIrIn$_5$ is much more itinerant than  known Ce-based HFs. We find that $1/T_1T$, subtracting that for LaIrIn$_5$, follows  a $(\frac{1}{T+\theta}) ^{\frac{3}{4}}$ variation with  $\theta$=8 K. We argue that this novel feature points to anisotropic,  due to a layered crystal structure,  spin fluctuations near a magnetic ordering.  The bulk superconductivity sets in at 0.40 K below which  the coherence peak is absent and $1/T_1$ follows a  $T^3$ variation, which suggests unconventional superconductivity with  line-node   gap.
\end{abstract}
\vspace*{5mm}
]
\narrowtext

The emergence of superconductivity  near a magnetic instability  in cerium (Ce)-based  heavy fermion (HF) compounds is one of the most intriguing phenomena in  strongly correlated electron systems.  Except for CeCu$_2$Si$_2$ which is superconducting  at ambient pressure      with $T_c$=0.65 K \cite{Steglich}, the superconductivity emerges near the quantum critical point (QCP) where the magnetic ordering is suppressed by large applied external pressure in CeIn$_3$ \cite{Mathur}, CeCu$_2$Ge$_2$ \cite{Jaccard}, CePd$_2$Si$_2$ \cite{Grosche} and CeRh$_2$Si$_2$ \cite{Movshovich}. In spite of efforts and progress,   however, knowledge about this class of superconductors is still limited because of difficult experimental conditions. The recently discovered new family of Ce-based heavy electron systems, CeMIn$_5$ (M=Rh, Ir)  with M=Ir being a superconductor already at ambient pressure \cite{Heeger,Petrovic0}, provides new  opportunities for studying the nature of the superconductivity in the vicinity of a magnetic instability,     the interplay between magnetic excitations and superconductivity, etc. In particular, CeIrIn$_5$ is suitable for studies using microscopic experimental probes that can be applied  more easily at ambient pressure.

CeMIn$_5$ (M=Rh, Ir) consists of alternating layers of CeIn$_3$   and MIn$_{2}$.
CeRhIn$_5$ is an antiferromagnet with $T_N$=3.8 K but becomes superconducting below $T_c$=2.1 K under pressures larger than 1.6 GPa \cite{Heeger}. In CeIrIn$_5$, the resistivity is already zero at ambient pressure below 1.2 K, but the Meissner effect and the jump in the  specific heat  are  found only at 0.4 K \cite{Petrovic0}. The electronic specific heat coefficient   $\gamma$ is found to be 750 mJ/mol K$^2$ \cite{Petrovic0}, which suggests a large mass enhancement. Recent de Haas-van Alphen Oscillation in CeIrIn$_5$ also reveals  a cyclotron mass that is $\sim$20 times larger than the band mass, consistent with the specific heat result \cite{Haga}.

In this Letter, we report a measurement using local probe, the $^{115}$In nuclear quadrupolar resonance (NQR) study in CeIrIn$_5$ down to 90 mK, at zero magnetic field. From the temperature ($T$) dependence of the nuclear spin lattice relaxation rate ($1/T_1$), we find that CeIrIn$_5$ is much more itinerant than  known Ce-compounds such as CeCu$_2$Si$_2$ \cite{Kitaoka}, and show that this compound is located near a magnetic ordering with anisotropic spin fluctuations due to the layered crystal structure. No anomaly was found at 1.2 K in the NQR quantities, but $1/T_{1}$ shows an abrupt decrease at 0.40 K below which the NQR intensity also decreases as does the ac susceptibility,  confirming a bulk superconductivity below  $T_{c}$=0.40 K. The lack of coherence peak  in $1/T_{1}$  just below  $T_{c}$=0.40 K followed by a power-law $T$-variation, $1/T_{1} \propto T^{3}$,   indicate that the superconductivity is of  unconventional type with an anisotropic gap. Our results show that CeIrIn$_5$ bares some resemblance to itinerant, quasi-two-dimensional high- $T_c$ copper oxides.

Single crystal of CeIrIn$_5$ was grown by the In-flux method
as in Ref. \cite{Heeger}. X-ray diffraction indicated that the compound is
single phase and forms in the primitive tetragonal HoCoGa$_5$ type structure. The resistivity already drops to zero at 1.2 K, which is in agreement with the reported property \cite{Petrovic0}.
The single crystal was crushed into powder to allow a maximal penetration of oscillating magnetic field, $H_1$. The measurements below 1.4 K were performed by using a $^{3}$He/$^{4}$He dilution refrigerator. A small $H_1$ was used to avoid possible heating by the RF pulse.  There are two inequivalent crystallographic sites of In in this compound, In(1) in the CeIn$_{3}$ plane and In(2) in the IrIn$_2$ plane.  Two sets of In NQR lines corresponding to these two sites were observed as shown in Fig. 1. The first set of the NQR  lines that are equally spaced is characterized by  $\nu_{Q}$=6.065$\pm$0.01MHz and the asymmetry parameter $\eta$=0. The second set of lines that are unequally spaced  was found at the positions centered at 33.700, 38.350, 52.185 and 71.432 MHz, respectively, which correspond to $\nu_{Q}$=18.175$\pm$0.01 MHz and $\eta$=0.462$\pm$0.001. Here $\nu_{Q}$ and $\eta$ are defined as $\nu_{Q}\equiv\nu_{z}=\frac{3}{2I(2I-1)h}e^2Q\frac{\partial V}{\partial z}$, and $\eta=|\nu_{x}-\nu_{y}|/\nu_{z}$,  with $Q$ being the nuclear quadrupolar moment, $I$=9/2 being the nuclear spin and $\frac{\partial V}{\partial \alpha} (\alpha=x, y, z)$ being the electric field gradient at the position of the nucleus \cite{Abragam}. The symmetric lines are assigned to the In(1) site and the asymmetric lines to the In(2) site, respectively, since  crystallographically In(1) site is axially symmetric but In(2) is not. This assignment is  consistent with the observation in CeIn$_3$ where $\eta$ is zero \cite{Kohori 0}. $\nu_{Q}$ for In(1) is  smaller, but  $\nu_{Q}$ for In(2) is larger by about 10\% than the respective values  in CeRhIn$_5$ \cite{Curro}. The transition lines are narrow with the full width at half maximum (FWHM) being $\sim$50 kHz, which indicates a good quality of the crystal. $1/T_1$ measurements  were done mostly on  the 1$\nu_{Q}$ ($\pm$1/2 $\leftrightarrow$$\pm$3/2) transition at low $T$ but at the 4$\nu_{Q}$($\pm$7/2 $\leftrightarrow$$\pm$9/2) transition at high $T$ for the In(1)-site.  The value of $1/T_{1}$ was obtained from the recovery of the nuclear magnetization following a single saturation pulse and an excellent  fitting  to the equations given by  MacLaughlin  \cite{Maclaughlin 1}. The value of $1/T_{1}$ measured at different transitions shows excellent agreement. 

Figure 2 shows $1/T_{1}$ as a function of $T$ in the temperature range of 0.09 K$\leq$$T$$\leq$100 K. We  discuss the normal-state property above $T_{c}$ first.  Remarkably,  $1/T_{1}$ shows strong $T$ dependence up to the highest temperature that we have measured, 100 K. This is to be compared to  other Ce compounds, such as CeCu$_2$Si$_2$ where $1/T_1$ becomes $T$-independent above $\sim$10 K \cite{Kitaoka}, which is assigned to be the Kondo temperature, $T_K$ below which the localized 4f moment is screened to produce a heavy quasiparticle state. This result   indicates that CeIrIn$_{5}$ is much more itinerant   than  other known Ce-compounds. In Fig. 3, we plot  $1/T_1T$ as a function of $T$.  The inset shows a $log-log$ plot that displays more clearly the behavior just above $T_c$. For comparison, we also show the data $1/T_1T$=0.81 Sec$^{-1}$K$^{-1}$ for LaIrIn$_5$ where no 4f spins are present. The  measurement for LaIrIn$_5$ was carried out at 4$\nu_Q$=23.77 MHz. It is seen that $1/T_1T$ of CeIrIn$_5$ is largely enhanced over that of LaIrIn$_5$ and increases strongly with decreasing $T$. Also note that a $T_1T=const.$ relation,  that would be expected from  the   Landau-Fermi liquid theory, is not obeyed. These aspects  indicate  that $1/T_{1}$ in  CeIrIn$_5$ is dominated by   itinerant spin fluctuations (SFs).   In fact, it was found more recently that substituting 40\% of Rh for Ir results in an antiferromagnetic (AF) ordering of the system at $T_N$=2.7 K \cite{Pagliuso}, which also suggests that CeIrIn$_5$ is a nearly AF metal.
In Fig. 4  we show $T_{1}T$ as a function of $T$. The open circles above 0.4 K correspond to $1/T_1T=1/T_1T$(CeIrIn$_5$)-0.81Sec$^{-1}$K$^{-1}$ (LaIrIn$_5$), which  represent the contribution due to the presence of Ce 4f spins alone.  Note  that this correction, by subtracting the relaxation of LaIrIn$_5$ which represents other relaxations including the In orbital contribution,  is negligibly small for 0.4 K$\leq T \leq$15 K.  As seen in the figure,  the data  can be fitted to a  relation of  $T_1T = C  (T+\theta)^{\frac{3}{4}}$
 with $\theta$=8 K  and C=4.75 msecK$^{\frac{1}{4}}$ (solid curve). 

 This unique $T$-dependence of $1/T_1T$ has never been observed in other HF compounds. We argue that this novel feature arises from {\it anisotropic} AF spin fluctuations, due to the {\it layered} crystal structure of CeIrIn$_{5}$. In d-, and also f-electron weak or nearly AF metals, many physical quantities    can be explained by 
the  self-consistent renormalization  theory for SFs  \cite{Moriya}. In this theory, it was shown that  the staggered susceptibility at   the AF wave vector $q=Q$, $\chi_{Q}$ (or the  squared magnetic correlation length) follows a Curie-Weiss (CW) variation above the Neel temperature in a weak AF magnet due to mode-mode coupling of SFs, namely, $\chi_{Q}\propto \frac{1}{T+\theta}$.   The value $|\theta|$  is   just the Neel temperature  ($T_N$) here.  In a nearly AF metal that does not order at finite temperature,  $\chi_{Q}$ is  shown to also obey a CW variation, while in this case  $\theta$   measures the closeness of a system to the magnetic ordering; $\theta$ decreases towards zero upon approaching the  ordering. 
Now, by assuming $\chi(Q+q,\omega)^{-1}=\chi_Q^{-1}+a q ^2 -i \frac{\omega}{\Gamma}$ it is shown that  $1/T_{1}T\propto\chi_{Q}$ for two-dimensional (2D) systems, but $1/T_{1}T\propto\chi_{Q}^{1/2}$  for 3D systems \cite{Moriya}.      Indeed, in many quasi-2D high-$T_c$ cuprates it is found $1/T_{1}T\propto \frac{1}{T+\theta}$   where the value of $\theta$ decreases upon approaching the magnetic ordering. For example, in the so-called overdoped compound TlSr$_{2}$CaCu$_{2}$O$_{6.8}$ which is far away from the magnetic ordering, $\theta$ is 235 K \cite{Zheng}. In less hole-doped system, La$_{2-x}$Sr$_{x}$CuO$_{4}$, $\theta$=120 K for $x$=0.24 while it decreases linearly with decreasing hole doping, reaching to $\theta$=20 K at $x$=0.075 \cite{Ohsugi}.   
In AF ordered 3D HF compounds, on the other hand, $1/T_{1}T\propto (\frac{1}{T+\theta})^{\frac{1}{2}}$ is well obeyed \cite{Kyogaku}. The predicted results by the 2D or 3D models are shown in Fig. 4.  As can be seen in the figure, although both models capture the low-$T$ behavior,   neither of them fits the data in the high-$T$ range. Let us now consider a situation where the dispersion of the SFs   is in-between 2D and 3D ones. If the SF dispersion in one direction is   flat, namely, the magnetic correlation length ($\xi$) is much  shorter  in one direction than in others, then by assuming $\chi(Q+q)^{-1}=\chi_Q^{-1}+a_1(q_x^2+q_y^2)+a_2q_z^4$ instead of   isotropic quadratic dispersion \cite{Lacroix}, it is     shown   that $1/T_{1}T \propto \chi_Q^{3/4}$. This anisotropic SF  model  explained the dynamical susceptibility in d-electron antiferromagnet  YMn$_2$, which orders   at $T_{N}$=110 K  but the ordering  can be suppressed either by applying external pressure or by substituting Sc for Y. In paramagnetic Y$_{0.97}$Sc$_{0.03}$Mn$_2$, inelastic neutron scattering measurement found that $\xi$ is shorter along the [001] direction ($\xi_{\perp}$=1.72 $\AA$) than that  along the [110] direction ($\xi_{\parallel}$=2.86 $\AA$), which is ascribed   to the geometrical  frustration of the magnetic interaction  \cite{Ballou}.  Indeed,  the same $T$-variation as found here, namely, $\frac{1}{T_1T}\propto (\frac{1}{T+\theta})^{\frac{3}{4}}$ was observed in paramagnetic YMn$_2$ under pressure \cite{Zheng1}.
On the above basis, we propose that the   $T$-variation of $1/T_1T$, that can be fitted to $(\frac{1}{T+\theta}) ^{\frac{3}{4}}$ in the entire $T$-range except near $T_c$,  is due to   anisotropic spin fluctuations  in CeIrIn$_{5}$. In fact, CeIrIn$_{5}$ has a layered crystal structure. Because of this 2D-like structure, a weaker magnetic correlation along the c-axis can be expected.  Further investigation by inelastic neutron scattering measurement would be interesting to confirm the SF dispersion in this compound. More systematic NQR/NMR study is also underway to see if the deviation of the low-$T$  data from the anisotropic SF curve points to any possible crossover to a different SF regime upon lowering $T$.  In any case, the small value of $\theta <$ 10 K indicates that CeIrIn$_{5}$ is located in close proximity  to the magnetic ordering.  Finally, we remark that the strong SFs near the magnetic ordering may also make an appreciable contribution to the huge specific heat.

Next, we discuss the superconducting (SC) state. First, as seen in Fig. 2, no anomaly was detected in $1/T_1$ around 1.2 K below which resistivity is zero. We also checked carefully the intensity and the linewidth of the NQR spectrum below 1.4 K; no anomaly is found when passing through 1.2 K.   However, $1/T_{1}$ decreases abruptly at $T=0.40$ K,  below which the NQR intensity decreases as does the ac susceptibility due to the Meissner effect. These results indicate that the bulk superconductivity sets in at 0.40 K, which is in good agreement with the specific heat measurement \cite{Petrovic0}. The  property of the SC state below 0.40 K is remarkable. Namely, 
 $1/T_{1}$ shows no coherence  peak just below $T_{c}$, and decreases in proportion to $\sim T^{3}$ upon  lowing $T$. This behavior is not compatible with  isotropic s-wave  gap. Rather, our result is qualitatively similar to that in other HF superconductors, such as CeCu$_2$Si$_2$ \cite{Kitaoka}, UBe$_{13}$ \cite{Maclaughlin}, etc \cite{Kohori,Kyogaku}, and  also  high-$T_c$ cuprate superconductors \cite{Asayama}, which indicates that the SC energy gap is anisotropic. 
 In terms of density of states (DOS), $T_{1s}$ in the SC state is expressed as $\frac{T_{1N}}{T_{1s}} = \frac{2}{k_BT}\int \int   (1+\frac{\Delta^2}{E E^{'}})N_{s}(E) N_{s}(E^{'}) f(E)(1-f(E^{'}))\delta(E-E^{'})dEdE^{'}$,
where $N_{s}(E)$ is the DOS in the SC state,
$f(E)$ is the Fermi function, $\Delta$ is the energy gap  and $C=1+\frac{\Delta^2}{E E^{'}}$ is called the coherence factor. In an isotropic s-wave superconductor, the divergence of $N_s$ at $E=\Delta$ results in  the coherence peak of $1/T_1$ just below $T_c$, and $1/T_1$ decreases as $exp(-\Delta/k_BT)$ at low $T$ because $N_s$=0 for $E<\Delta$. By contrary, an anisotropic gap generally reduces the divergence of $N_s$  and produces a finite DOS at low energy. For example, if we assume a line-node  gap of
$\Delta(\phi)=\Delta_{0}\cos \phi$,  then by integrating $C$ and $N_s=\frac{N_0 E}{\sqrt{E^{2}-\Delta(\phi)^{2}}}$ over $\phi$ we obtain $C$=1,  $N_s=\frac{\pi}{2}\frac{N_{0}E}{\Delta_0}$ for $E\leq\Delta_0$ and $N_s=\frac{N_{0}E}{\Delta_0}sin^{-1}(\frac{\Delta_0}{E})$ for $E\geq\Delta_0$. The finite value of $N_s$ at $E=\Delta_0$  removes the coherence peak and the $E$-linear DOS below $\Delta_0$ gives rise to a $T^3$ variation of $1/T_1$ at low $T$. 
   The curve below $T_c$ in Fig. 2 depicts the    calculated result assuming the above model with $2\Delta_0=5.0 k_BT_c$ and a BCS $T$-dependence for $\Delta_0$.       
This  gap amplitude $\Delta_{0}$ is about the same as that  for CeCu$_2$Si$_2$ to which should the same model be applied. It is, however,  substantially  smaller  than that in some uranium (U)-based HF superconductors where $2\Delta_0$ would reach $\sim 10k_BT_c$ for the same gap function\cite{Maclaughlin,Kohori}, which  may  be related to its  proximity to a magnetic instability of the present compound. In fact, a recent study
 found a reduced $\Delta_{0}$ in  Ce$_{0.99}$Cu$_{2.02}$Si$_2$  
\cite{Kawasaki}, which is believed to be located closer to  QCP than a stochiometric compound \cite{Geibel}. Applying external pressure increases $\Delta_{0}$ \cite{Kawasaki}. In CeIrIn$_{5}$,  applying  pressure increases $T_{c}$ \cite{Petrovic0}. Further investigations by NQR under pressure are in progress in order to reveal how the gap amplitude evolves with pressure, and the intimate relation between the superconductivity and the magnetic exitations.

In summary, we find from NQR $1/T_1$ measurement that  the new  heavy fermion (HF) compound CeIrIn$_5$ is much more itinerant than known Ce-based HFs. We further find that $1/T_1T$, subtracting that for LaIrIn$_5$,  follows  a $(\frac{1}{T+\theta}) ^{\frac{3}{4}}$ variation with a small $\theta$=8 K, and interpret it as arising from anisotropic, due to the layered crystal structure, spin fluctuations near a magnetic ordering.  Below $T_c$=0.40 K, $1/T_1$ decreases in proportion to $T^3$, which indicates unconventional superconductivity with an anisotropic energy gap.  Our results show that the superconducting state of this new compound  is reminiscent of those in known HF  and high-$T_c$ cuprate superconductors, but the normal-state magnetic excitation lies   in-between those for the quasi-2D high-$T_{c}$ cuprates and  the known 3D, more localized HFs. Therefore, the knowledge about this compound may also be useful for understanding the physics of both high-$T_{c}$ cuprate and traditional HF superconductors.

We thank J. Sarrao, C. Petrovic and  J. D. Thompson for discussing with us about their results on pure and Rh-substituted CeIrIn$_5$ prior to publication, T. Takeuchi, Y. Kawasaki and K. Ishida for useful discussions, and
partial support by a Grant-in-Aid for Scientific Research No.  10CE2004  from
the Japanese Ministry of Education, Science, Sports and Culture.

\begin{figure}[htbp]
\caption[]{Two sets of $^{115}$In NQR lines observed at $T$=4.2 K corresponding to  In(1) site in the CeIn$_{3}$ plane and  In(2) site in the IrIn$_{2}$ plane.  The small peaks to the left of each line of In(2) are from the less abundant isotope $^{113}$In. } 
\end{figure}

\begin{figure}[htbp]
\caption[]{$T$ dependence of the $^{115}$In nuclear spin-lattice relaxation rate. The solid curve is a calculation by assuming a line-node gap $\Delta(\phi)=\Delta_{0}\cos \phi$ with $\Delta_{0}=2.5k_{B}T_{c}$ (see text for detail). }
\end{figure}

\begin{figure}[htbp]
\caption[]{The $1/T_1T$ of $^{115}$In plotted as a function of $T$.  The inset shows a $log-log$ plot that emphasizes the low-$T$ behavior just above $T_c$. Also shown are data for LaIrIn$_5$. A strong increase of $1/T_1T$ with lowing $T$ points to the antiferromagnetic spin correlation.} 
\end{figure}

\begin{figure}[htbp]
\caption[]{$T_1T$ plotted as a function of $T$. Solid circles are the raw data of CeIrIn$_5$, while open circles correspond to the normal-state relaxation rate  subtracting that of LaIrIn$_5$. The broken line, solid and dotted curves are the $T$-variations of $T_{1}T\propto T+\theta $  (2D SFs), $T_{1}T\propto (T+\theta)^{3/4}$ (anisotropic SFs) and  $T_{1}T\propto (T+\theta)^{1/2}$ (3D SFs), respectively.}
\end{figure}
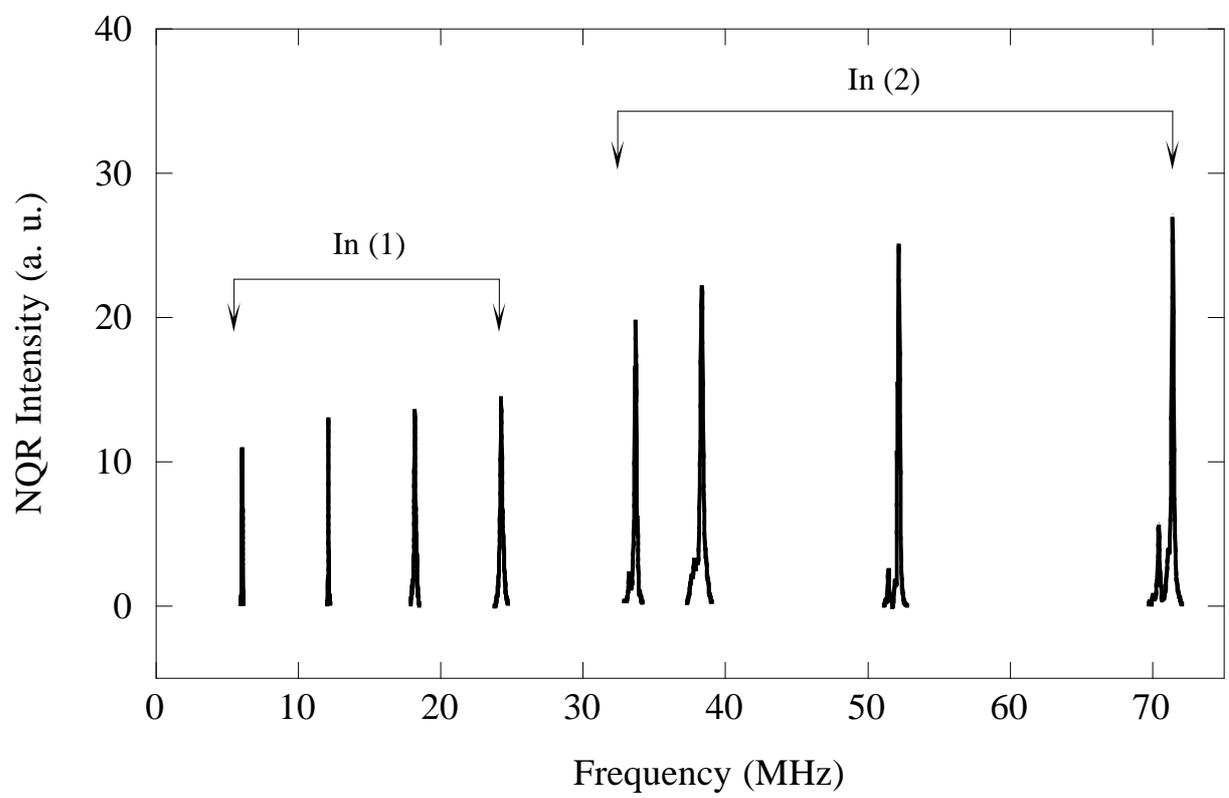
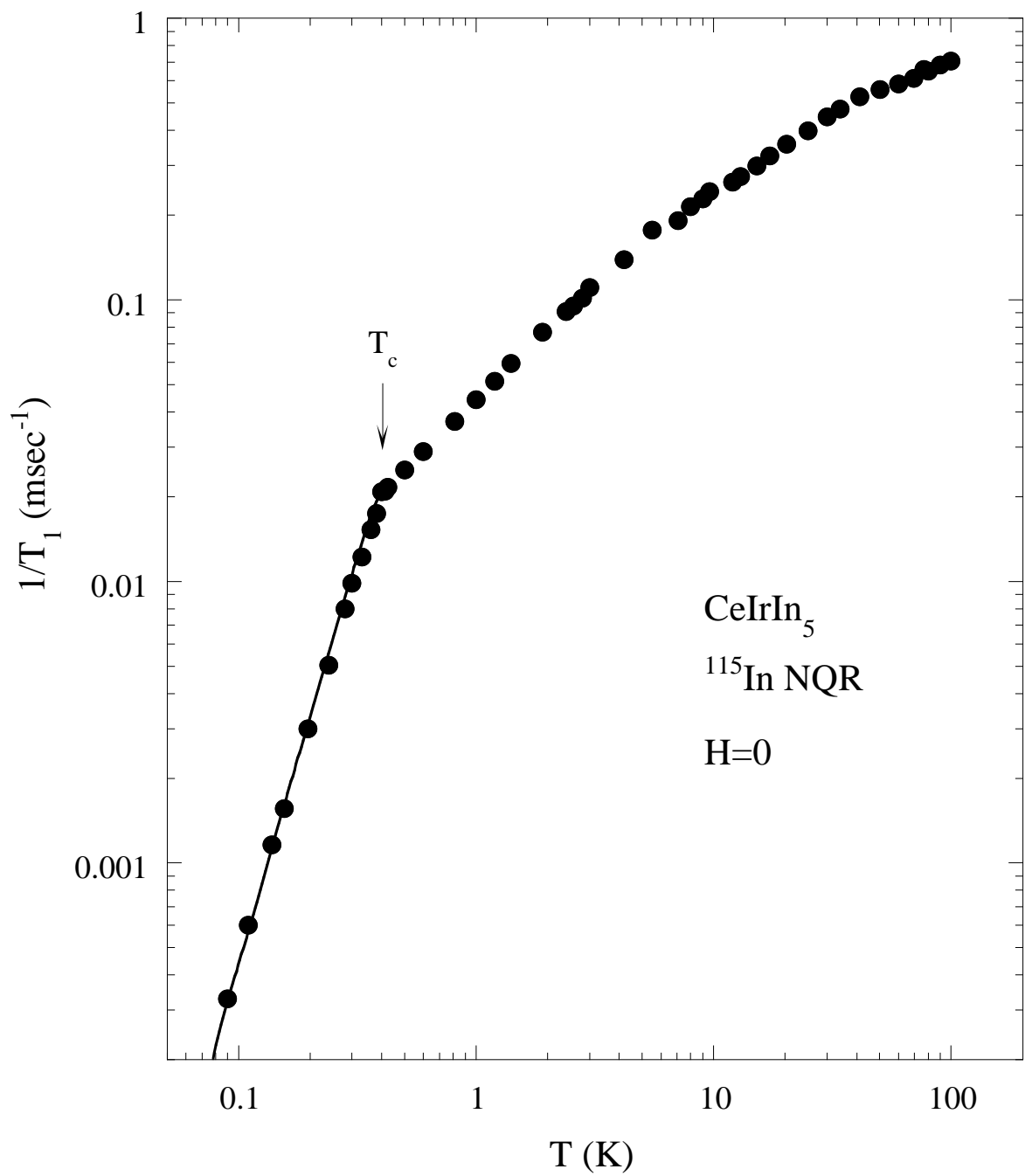
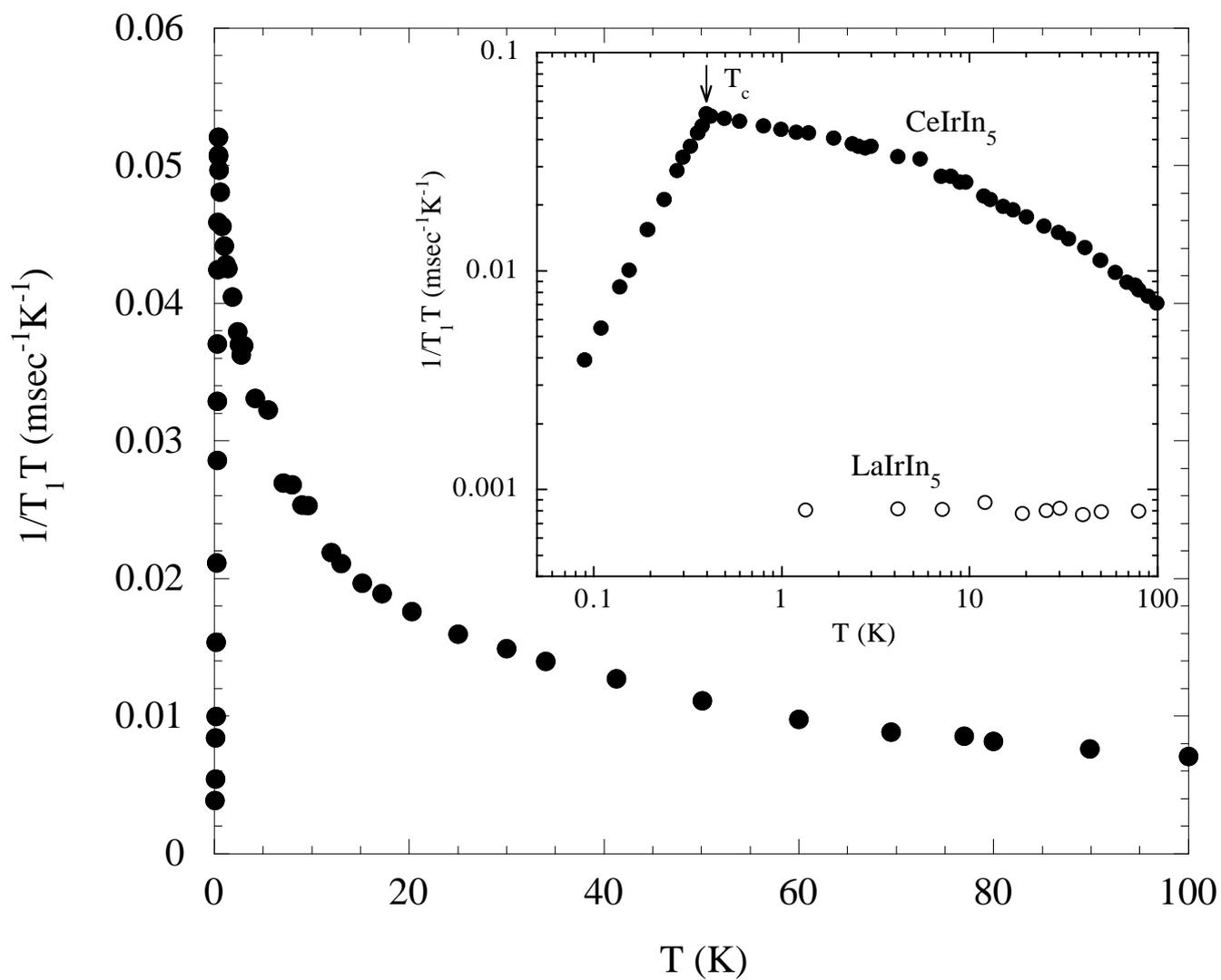
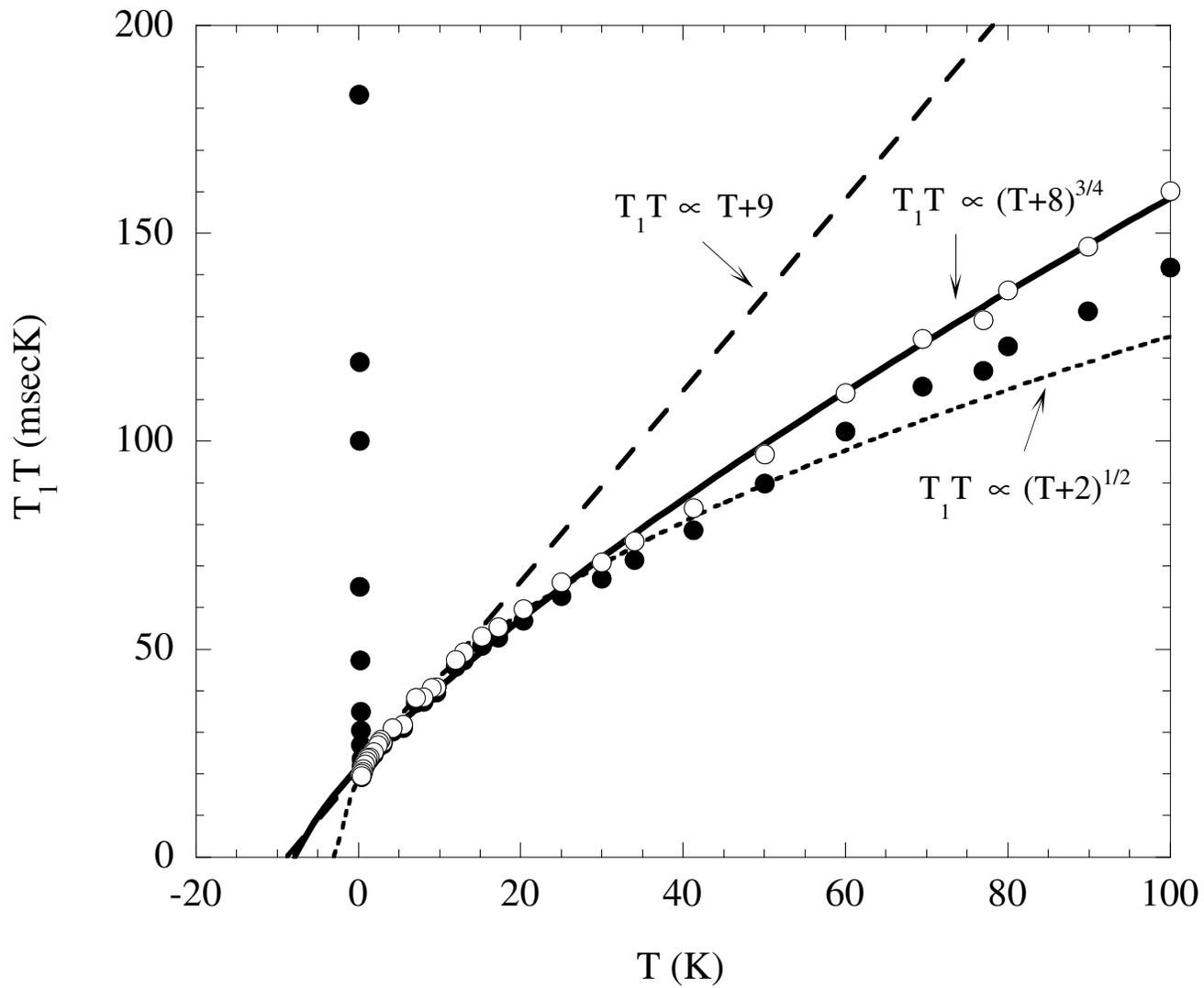
\end{document}